# Accurate single-shot measurement technique for the spectral distribution of GeV electron beams from a laser wakefield accelerator

C. I. Hojbota[1,2], Hyung Taek Kim[1,3*], Jung Hun Shin[1], C. Aniculaesei[1], B. S. Rao[1,4], Chang Hee Nam[1,2**]

*[1]Center for Relativistic Laser Science (CoReLS), IBS, Gwangju 61005, Korea*

*[2]Department of Physics and Photon Science, GIST, Gwangju 61005, Korea*

*[3]Advanced Photonics Research Institute, GIST, Gwangju 61005, Korea*

*[4]Laser Plasma Division, Raja Ramanna Centre for Advanced Technology, HBNI, Indore452013, India*

*[htkim@gist.ac.kr](mailto:htkim@gist.ac.kr), **[chnam@gist.ac.kr](mailto:chnam@gist.ac.kr)*

We present a technique, based on a dipole magnet spectrometer containing multiple scintillation screens, to accurately characterize the spectral distribution of a GeV electron beam generated by laser wakefield acceleration (LWFA). An optimization algorithm along with a numerical code was developed for trajectory tracing and reconstructing the electron beam angle, divergence, and energy spectrum with a single-shot measurement. The code was validated by comparing the results with the Monte-Carlo simulation of electron beam trajectories. We applied the method to analyze data obtained from laser wakefield acceleration experiments performed using a multi-Petawatt laser to accelerate electron beams to multi-GeV energy. Our technique offers improved accuracy to faithfully characterize electron beams with non-negligible shot-to-shot beam pointing fluctuations, particularly in the state-of-the-art multi-GeV LWFA experiments performed to push the energy frontier.

## 1. Introduction

In laser wakefield acceleration (LWFA)[1], an ultrashort laser pulse drives a nonlinear plasma wave, which in turn can trap and accelerate electrons to energies up to multi-GeV [2,3]. Progress in the past decades led to a dramatic increase in the energy and quality of accelerated electron beams. Nowadays, LWFA can produce electron beams with an energy spread as low as 1% and shot to shot stability below 10 % [4]. Recent development of novel techniques, such as multi-staging[5], tailoring of the density profile of a target[6], or plasma guiding techniques[7], offers the opportunities to achieve high-energy high-quality stable electron beams in near future, potentially complementing established technologies of electron accelerators. Such compact electron accelerators are the focus of intensive research due to their wide range of applications: electron diffraction, generation of coherent X-rays through the operation of free electron lasers, production of intense gamma-rays from Compton backscattering and exploration of fundamental QED processes in all-optical setups.

A precise and accurate method to measure the energy spectrum of electron beams in LWFA is essential since energy is one of the most crucial parameters characterizing accelerated electrons. Shot-to-shot variations of electron beam parameters, such as energy, energy spread, emittance, divergence and beam pointing, make the precise measurement of these parameters difficult[8]. In particular, experiments with multi-PW lasers for obtaining high energy electrons could suffer from the fluctuation of electron energy and beam pointing due to the intrinsic complexity of a large laser system and the instabilities grown during the propagation of an intense laser pulse through a plasma medium. The pointing fluctuations of the electron beam can be an unavoidable source of non-neglegible errors in the energy measurement using dipole magnets. The challenge of finding an accurate calibration technique for LWFA electrons stimulated past research: some studies used magnets with custom shapes[9,10] or multiple magnets[11] to accommodate a wide energy range; other studies accounted for the the change in trajectories due to pointing fluctuations by using a beam profile monitor[12] or observing secondary x-rays[13], and used multiple fiducial wires in order to trace back the incoming beam angle[14,15]. Large pointing stability still affects the final estimations particularly at high energy (beyond 1 GeV). Thus, a method for faithfully reconstructing the energy distribution of an electron beam is crucial for accurate characterization of beam parameters.

In this work we present an energy characterization method, based on a magnetic dipole spectrometer containing multiple scintillating (Lanex) screens and an in-house developed numerical code, for acurately reconstructing GeV-class electron beam parameters. We validated the method for reconstructing the trajectories of electrons for various values of pointing and energy by comparing it with GEANT4 and gyroradius calculations. The method was then applied to characterize multi-GeV electron beams generated from LWFA experiments performed at the 4 PW beamline at the Center for Relativistic Laser Science (CoReLS).

The content in the article is structured as follows. After the introduction in Section 1, we discuss the reconstruction method in Section 2. We describe the algorithm used and compare the errors in the deflection obtained with the existing code (GEANT4) and an analytic estimate (gyroradius). We then estimate the reconstruction errors for several simulated trajectories with different energies and angles and discuss the effects of miscalibration. Section 3 describes in depth the experimental setup used for measuring the angle and energy of electron beams. In Section 4 we present results from LWFA experiments performed at CoReLS and apply our calibration method to obtain the energy calibration of electron beams. We end this article with conclusions and discussions on further improvement of this method.

## 2. Reconstruction methodology

### 2.1 Basic algorithm for reconstructing beam energy and angle

In order to accurately reconstruct the electron beam energy and angle, we developed a code, ENCALMS - Electron eNergy CALibration for Magnet Spectrometers. The code has a user interface and works within a realistic range of parameters for the electron beam from LWFA. It can take into accout parameters of the position of the beam on the 1st Lanex screen (L1) before a dipole magnet and the positions of energy peaks on the 2nd and the 3rd Lanex screens (L2 and L3) after the magnet (more details about the electron spectrometer setup will be given in the experimental section). It then provides output paramers viz., the optimal energy and angle of the beam that produced the measured parameters. The code is separated into two parts: propagation through a given setup and the reconstruction of the electron beam trajectory (energy and angle).

The particle propagation part is divided into two steps: ray tracing and propagation through a magnetic field. The division of the propagation into these two was chosen for efficient computation by reducing the required computational power, as the ray tracing is computationally less intensive; the whole propagation is done in 2D in the dispersion plane of the electron beam. The setup for ENCALMS can be customized by adding several screens and one or more magnets. A 4th order Runge-Kutta (Nystrom) algorithm is used for the 2D propagation of the relativistic electron through a custom magnetic field. This is a standard method that has been used, for example, in GEANT4[16] , or for tracking particles at the ATLAS detector[17] . The field is loaded on a grid and at each instance of a step we perform a 2D linear interpolation. The ray tracing part calculates the position of a particle in the propagation plane, given the initial starting point, angle and the longitudinal distance. The method is then used to find the intersection with objects such as screens, which are parametrized as lines on a plane.

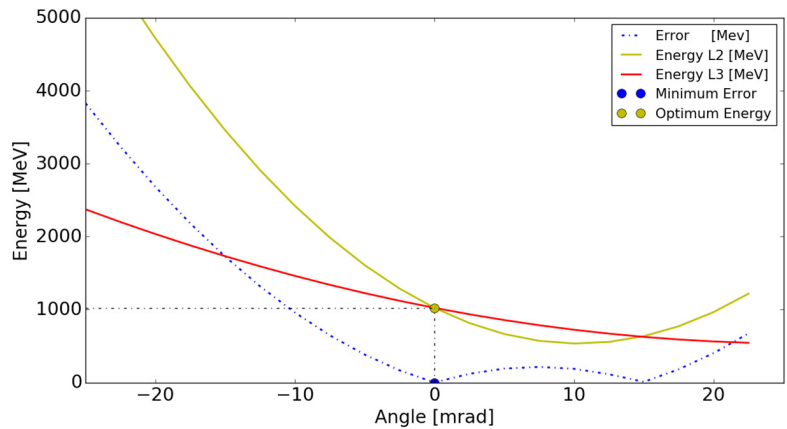

Figure 1 Illustration of error minimization. The red and the yellow lines represent the combination of angle and energy of electrons that intersect L2 and L3 respectively, at given points. The point where the error is minimized corresponds to the optimal combination of angle and energy. The simulation was performed for a particle of 1 GeV energy and direction perpendicular to the magnet edge.

The second part of the program estimates the error in the reconstructed energy by comparing the position at which multiple trajectories hit the detector screens with the centroid of the beam measured during experiments. For each position on a screen outside of the magnet, there are multiple combinations of angle and energy that go through the same point, as a consequence of magnetic deflection and focusing. Our algorithm calculates the intersection of these angle-energy functions for L2 and L3 in our setup and the intersection point corresponds to the angle-energy combination that passes through both screens. The angle-energy functions can be obtained by simulating trajectories with multiple initial angles and energies, then an interpolation is performed to obtain the accurate energy through a given point in L2 and L3. We illustrate this concept in Figure 1, where we can observe that the intersection point of two functions (one for L2 and another for L3) correspond to the energy and the angle of an incoming particle. Although there are 2 intersection points, we should consider that the second one corresponds to an unphysical larger incoming angle; such a case would be obviously sorted out by measuring the beam position on L1. Since the range of possible angles is limited and by considering restrictions of the physical measurement, we can search for this unique point within a reasonable interval over the beam cross-section, monitored at L1 in front of the dipole magnet. In this way, by finding the point that minimizes the error between the two angle-energy functions, we can reduce the uncertainty in the measured energy for fluctuating beam angle and distribution.

## 2.2 Calibration

In order to assess the accuracy of our method, we first calibrate it with existing codes. For this, we use two references: simulations performed with GEANT4 and comparison with an analytic formula for the gyroradius. These tests should make sure that the code works within a range of parameters suitable for experiments, and the errors are small compared to the uncertainty in measurement.

### 2.2.1 Comparison of GEANT4 – ENCALMS

In Figure 2 we show a graphical comparison of electron trajectory from GEANT4 and ENCALMS. To examine the errors, we chose a setup consisting of three screens perpendicular to the axis of incidence, placed at 50, 100 and 140 cm from the source, respectively. The magnet with a size of $30\times10$ cm$^2$, is placed 70 cm from the source, centered on the axis of propagation and it is initialized with a uniform magnetic field of 1 T.

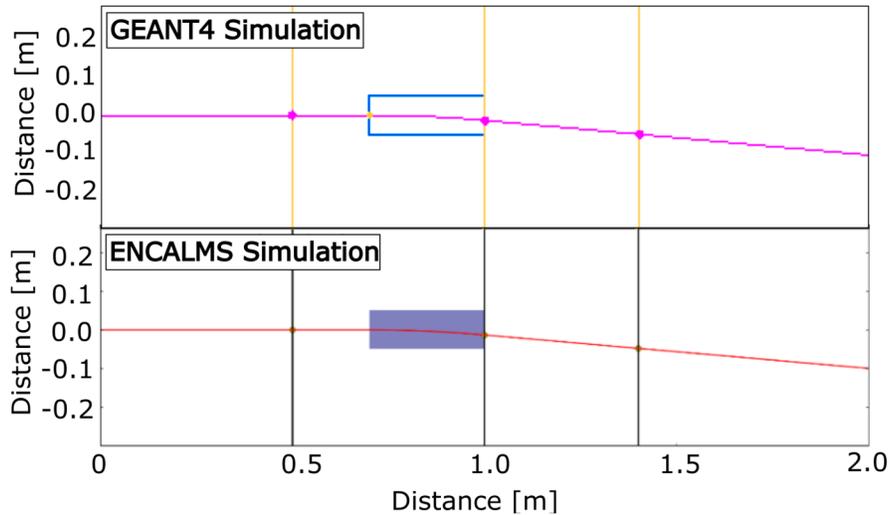

Figure 2 Electron trajectory simulated with GEANT4 (pink) and with ENCALMS (red). The magnet (blue in both cases) bends the trajectory of an electron, and the electron intersects the three screens (yellow – GEANT4 and black – ENCALMS).

We performed a series of simulations for different angles and energies to compare the transverse deflection of the electron beam with different methods. We show the difference between GEANT4 and ENCALMS in Figure 3. We plot the absolute value of the deflection error, in percentage (Figure 3 a.), as a function of the incoming particle energy. The error is defined as the difference in deviations along the transverse axis, at the exit of the magnet, relative to the deflection from ENCALMS. We note that the difference in the deviation is around 0.1% between the results obtained in ENCALMS and those from GEANT4. The error is small, but the absolute value of the energy miscalibration increases with the incident energy, as it is illustrated in Figure 3 b. For all purposes, an error of 14 MeV for an electron beam of 10 GeV is much smaller than the energy spread of current electron beams from LWFA. Although the error is small, the source of such an error can be the algorithm used for solving the equation of motion. The error could come from the fact that the RK method implemented in GEANT4 has a convergence part, where the interval is successively divided until the propagation error becomes smaller than an established limit. Additionally, we note that in our case, such optimization would not be necessary since the error in the propagation algorithm would be smaller than the one from the physical resolution of the dipole magnet.

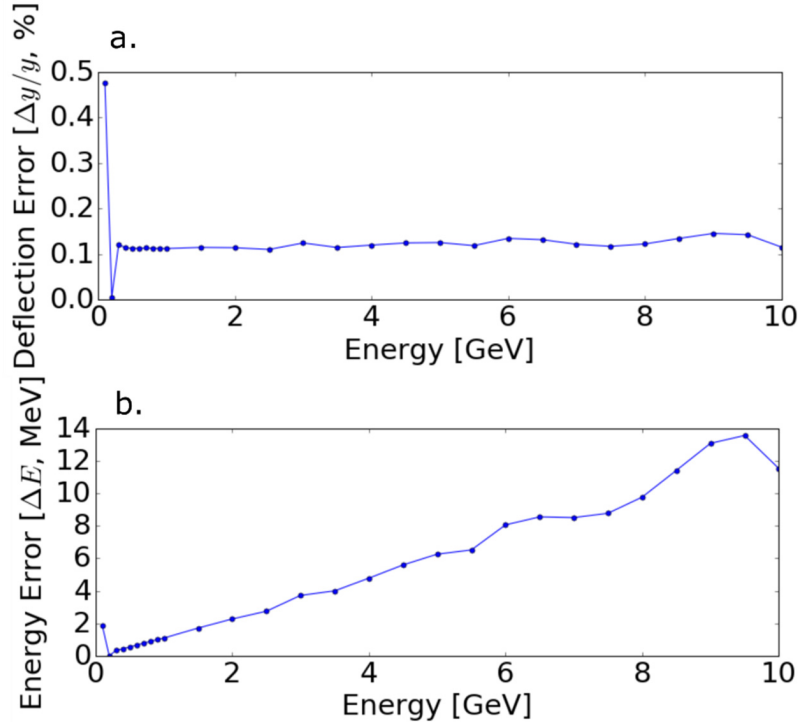

Figure 3 a. Propagation error. Comparison between trajectory deflections from GEANT4 and ENCALMS for the energy range 0.1 – 10 GeV. b. Absolute error in the energy calibration for the simulations performed with GEANT4. The error is given by the difference between the particle energy simulated with ENCALMS and the energy corresponding to the position of the deflected particle in GEANT4.

### 2.2.2. Comparison with a gyroradius

Since the errors between GEANT4 and ENCALMS can be a source of difference in the numerical algorithms, we compare the results of the trajectories obtained with the analytical formula for the gyroradius ($r_g = \frac{\gamma m v_\perp}{|q|B}$, where $\gamma$ is the relativistic factor of an electron beam, $v_\perp \cong c$ is the velocity transverse to the magnetic field direction, m electron mass, q=-1 electron charge and B the magetic field). We implemented the formula in ENCALMS and observed the deviation along the transverse axis, comparing it with the numerical propagation. We show the results in Figure 4, where we plot this error for a range of energies, and zero angle of incidence (Figure 4 a.). We note that the error is on the order of $10^{-3}$ and converges again to a value about 0.1 %. While the error in the energy calibration also increases with energy, as the comparisons with GEANT4, the error remains smaller than 10 MeV even for incident energies of 10 GeV (see Figure 4 b.). Thus, our code agrees reasonably well also with the analytical estimations for the trajectories.

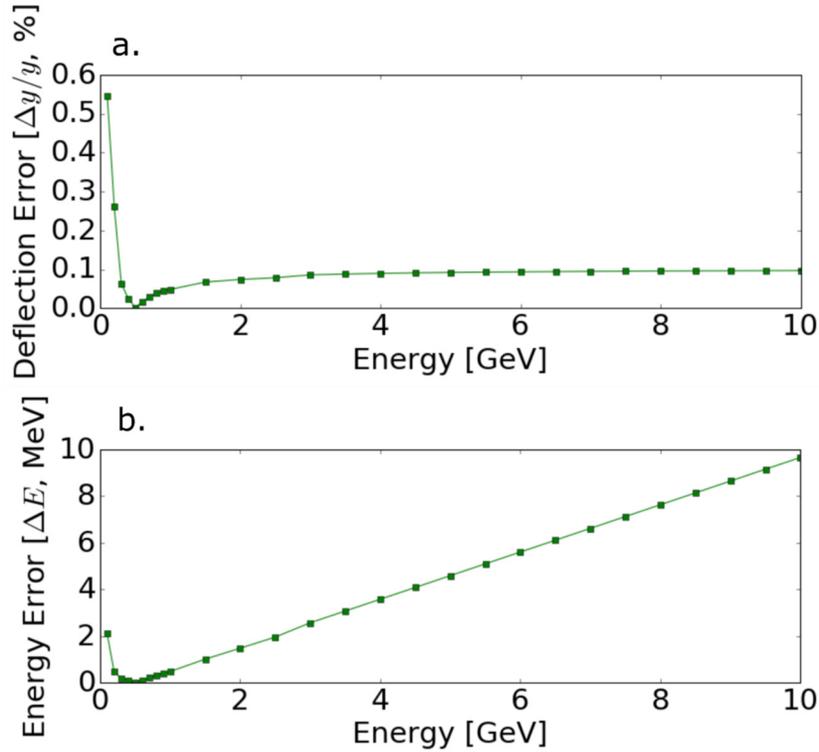

Figure 4 a. Comparison between the trajectory deflection simulated with ENCALMS and the ones from the analytical formula for the gyroradius, for the energy range 0.1 - 10 GeV; b. Absolute error in the energy calibration obtained by comparing with the gyroradius calculation. We computed the error by taking the difference between the energy of the particle simulated using the ENCALMS code and the energy corresponding to the position of the deflected particle using the analytical estimation of the gyroradius.

### 2.3 Reconstruction

We used the algorithm to reconstruct the angle and energy for a set of beam trajectories. We simulated the propagation of electrons through a setup consisting of three Lanex screens (L1, L2 and L3) placed at 1.0 , 1.5 and 2.0 m along the propagation direction, and a $20 \times 40$ cm$^2$ magnet with a uniform field of B=1 T, placed before L2. We performed the simulation for the beam having energies E={0.5, 1, 2.5, 5, 7.5, 10} GeV and incidence angles $\theta$ ={±50, ±30, ±10, ±5, ±1} mrad. To reconstruct the trajectories, we scanned a range of 20 energies [0.5E, 2E] and 10 angles [$\theta$ - abs($\theta$), $\theta$ +abs($\theta$)], and gave the intersection position of the electron trajectories on the screens.

Figure 5 presents the results obtained from reconstruction. The error in energy in Figure 5 (a) and (b) ranges of 0 - 1×10$^{-2}$ for positive incidence angle, and 0 - 1×10$^{-2}$ for negative angle. The error of the estimated angle, presented in (c) for positive and (d) for negative deflections, is a small value below 10$^{-3}$. This amount of error would be smaller than the resolution of a Lanex screen placed at 1 m from the target. For example, for a screen required to span a range of 1 - 100 mrad and a CCD sensor with a resolution of 1024 pixels, the minimum angular resolution would be 0.1 mrad, which corresponds to $10^{-3}$ for an angle of 100 mrad. Thus our algorithm can give a better accuracy than the instrumental resolution. Observing the errors in Figure 5 (a) - (d), we can conclude the algorithm is accurate enough for our purpose and the error can be further reduced by increasing the accuracy of the scanning interval.

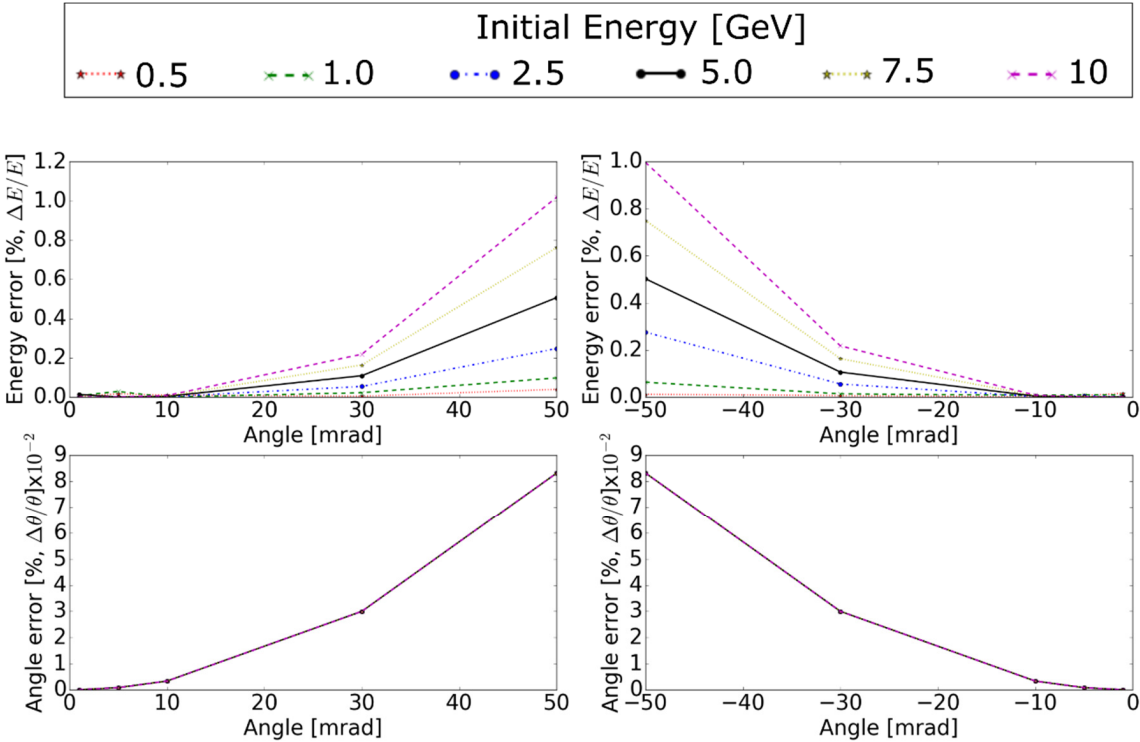

Figure 5 Reconstructed trajectories for the energy 0.5 - 10 GeV and angles from -50 mrad to 50 mrad. We show here the error in the reconstructed energies for positive (a) and negative (b) incidence angles. The error in the reconstructed angle is shown for initially positive (c) and negative (d) incoming angles.

### 3.4. Energy miscalibration

The most important aspect of using our method is accounting for the angle of an incident beam. Calibration could be done, of course, by assuming a 0 deg incidence angle (i.e. without correcting for the beam pointing jitter during experiments). The reconstruction error, in this case, would be high and the spectrometer cannot be accurately used for determining the electron beam energy. In order to evaluate quantitatively the error induced by the incidence angle, we present reconstruction results in different conditions. We simulate an electron trajectory for a range of angles θ = [-30 … 30] mrad and energies E = [0.5 … 10] GeV, and then we attempt to reconstruct the initial energy by erroneously assuming that the beam propagates straight along the laser axis. In Figure 6 we plot the error between the initial energy (E) and the reconstructed energy (E`) on a logarithmic scale. As we can observe in the figure, the error varies significantly across the energy–angle landscape. The considerably small errors laid only at low incidence angle and low electron energy.

As the angle increases, so does the error in the estimated energy. For low energy (0.5 GeV), the error will range from a few percents at θ = 0 mrad up to 90% for θ=30 mrad. As the energy increases, the error due to a larger deviation angle grows drastically, leading to errors of roughly 2 orders of magnitude the initial energy. Such an error is a natural consequence of the deviated beam trajectory by incidence angle and it can indicate that some measurements could not be performed properly. In addition to the variation in the relative error, we can observe an asymmetry between positive and negative angles. For positive angles, the electron trajectories will be bent by the magnet closer to the laser axis, which means they will hit the screen where 0-degrees electrons should have higher energy, and the scale is finer. Due to this reason, a transverse displacement leads to a higher energy range for positive angles, than for negative ones. This error evalutation shows that ignoring the initial angle leads to an entirely unreliable calibration.

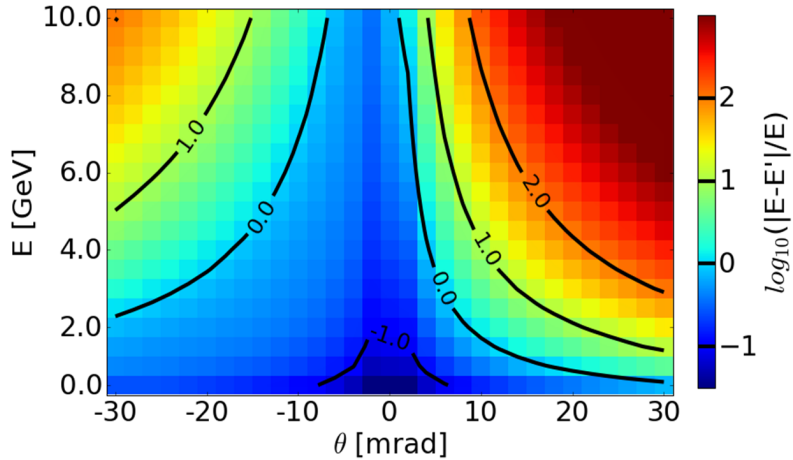

Figure 6 Relative calibration error for a particle assumed to be at zero incidence angle. The relative difference between the initial energy (E) and the reconstructed energy (E`) is shown as a function of incidence angle and energy.

### 3. **Experimental Setup**

The experiments to accelerate electrons to multi-GeV energy were performed using a multi-PW laser beamline at CoReLS[18–20] . For the acceleration process, a laser beam of 40-J energy with 23-fs pulse duration was focused onto a He gas cell target. The temporal pulse duration was measured with a self-referenced spectral interferometry setup[21] that is part of a closed-loop feedback system allowing to manipulate the spectral phase of the PW laser. Combined with an acousto-optic programmable dispersive filter (AOPDF), the temporal envelope and spectral phase of a PW laser pulse can be optimized to increase the energy of electron beams[22,23]. A spherical mirror with a focal length of 15 m (f-number = 50) was used to focus the beam to a spot of 65 μm in full-width at half-maximum. Before the final focusing, a deformable mirror[24] was used to flatten the wavefront, which is crucial for obtaining a good quality of the focal spot. The experimental parameters were chosen such that the beam size, the normalized vector potential and the plasma density could match[25] . The plasma density was optimized to be around $10^{18}$ cm$^{-3}$ for the most stable electron beams, and the length of the gas cell was adjusted to obtain maximum energy. The design of the gas cell was based on simulations (ANSYS fluent) and generated a uniform plasma density profile. More details about the gas target can be found in[26] .

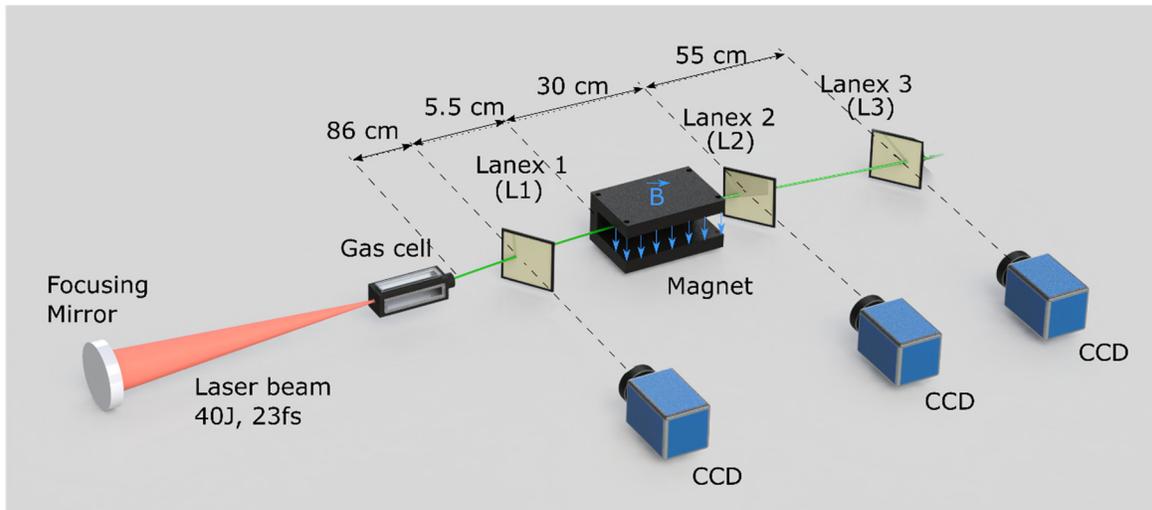

Figure 7 Experimental setup for laser wakefield acceleration. A spherical mirror focuses the beam onto a He gas cell. After the acceleration, the electron beam (green) propagates through a dipole magnet spectrometer for the measurement of energy spectrum and its image is recorded on three scintillating screens (Lanex) coupled to CCDs.

The energy spectrum of an accelerated electron beam was measured with a dipole magnet spectrometer, as shown in Figure 7. For the measurement, the partially depleted laser beam, co-propagating with accelerated electron beam, was blocked by Al foils installed in front of L1. The electron beam, after passing through L1 and the foil, enters the spectrometer, which consists of three scintillating screens and a C-type permanent dipole magnet based on NdFeB. The magnet is a 30 cm long device with 8.8 cm pole width and 3 cm pole gap. A longitudinal profile of the magnetic field is presented in Figure 8; the fringe field continues for several centimeters past the pole edges, and the maximum magnetic field strength corresponds to B = 1.33 T. The transverse field disperses an entering electron beam according to the particle energy, and the length and strength of the field. These are chosen in such a way as to resolve electron energy up to 3 GeV at the exit of the magnet. The resolution dE/E increases with distance from the magnet, allowing us to measure spectrum up to 7 GeV at 55 cm away from the spectrometer exit.

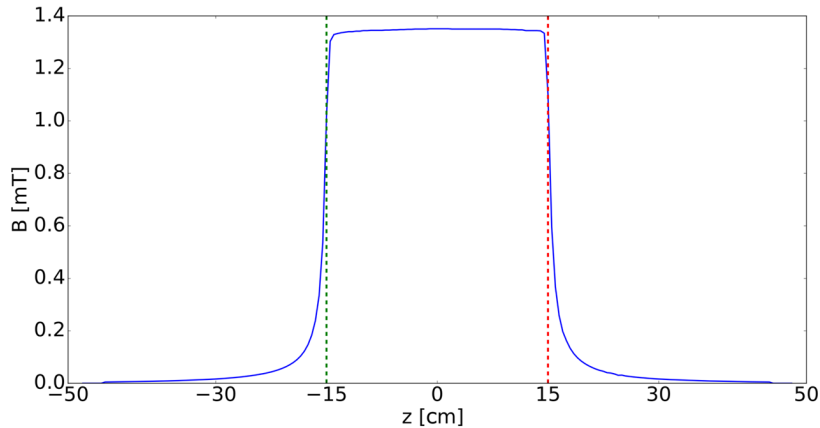

Figure 8 Magnetic field profile (blue line) and the magnet edges (dashed green line) along the laser propagation axis.

The electron beam was monitored using three Gd$_2$O$_2$S:Tb screens (Lanex Back), denoted L1, L2, and L3. The first screen L1 was placed before the magnet and imaged onto a 14-bit CCD sensor (PCO Pixelfly) using a zoom lens (Tamron AF 28-300mm). The purpose of the screen was to measure the divergence of an electron beam and the incidence angle before the magnet. This measurement before the dipole magnet is of crucial importance: without a proper evaluation of the incidence angle, the calibration would be erroneous, since it introduces an additional deflection in the dispersion plane. The second screen L2 was installed right at the exit of the magnet and the dispersed electron beam was imaged onto a 16-bit back-illuminated EMCCD sensor (Andor iXon Ultra888) using a zoom lens (Samyang Reflex 300mm), placed at 125.7 cm from the screen. For low energy electrons below 2 GeV, the electron energy can be recalibrated by measuring L2; however, if the energy of electrons is high, the energy resolution becomes poor (dE/E≈6% at E=2GeV). To accurately measure high energy electrons we needed one more screen after the magnet. Therefore we installed L3 at 53.0 cm away from the magnet exit and rotated it to a 45 degrees angle, perpendicularly to the electron dispersion plane. The screen is imaged onto an EMCCD, using a zoom lens (Samyang Polar 35mm f1:1.4) placed at 110 cm away. The Lanex films were charge calibrated using a RF-driven 20pC electron beamline at KAERI[27] . Our algorithm takes into account all 3 screens, reducing the error in measured energy peaks.

## 4. Experimental Results

We present here LWFA experimental results and their energy-angle calibration. Figure 9 and Figure 10 show images on the three screens and calibrated electron spectra for five shots (1-5), respectively. The laser energy on target and the pulse duration were about 40J and 23 fs, respectively. The medium used for acceleration was He with 3% Ne, the inlet pressure of the gas cell was 10 bars, and the acceleration distance was set to 5 cm in order to fit the dephasing length. The electron energy obtained in these shots was in the ragne of 2 - 4 GeV, which are enough to show the capability of our calibration method.

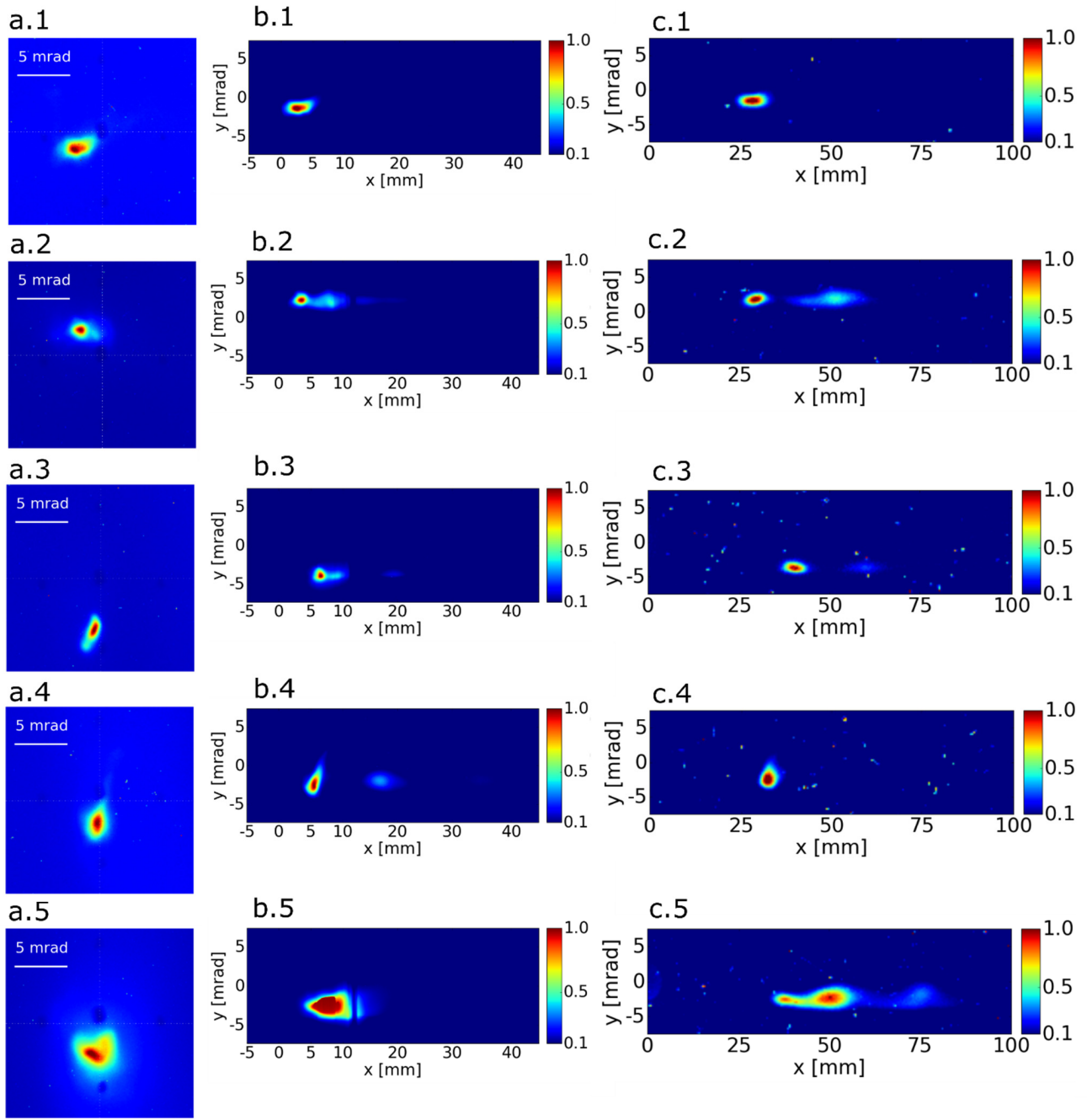

Figure 9- Raw data of accelerated electron beams. The figure shows the acquired images generated for 5 shots (1-5) as the beam passes through L1 (a), L2 (b) and L3 (c). The color represents the intensity of the optical emission from the Lanex screens on a linear scale, uncalibrated but normalized to the peak intensity of each image.

In Figure 9 we show images of the emission generated by five electron beams on the scintillating screens. We used the setup described in Figure 7. The transverse beam profile on L1 is shown in the first row of Figure 9 (a.1-a.5), and it has a peak-to-valley divergence in the range of 1 - 5 mrad. The darker faded dots along the transverse plane axes (horizontal and vertical) serve as references for the imaging setup, the markings are written on L1 and spaced at 5 mm distance each. After propagation through the magnet, the dispersed beams were imaged on L2 (b.1-b.2) and L3 (c.1-c.5). Since the third screen was placed farther away from the magnet, the angular distribution of the beam has an elongated profile, as seen in Figure 9 (c). From the measured data on the Lanex screens the input parameters for the calibration with ENCALMS were collected.

Calibration was performed for the energy range 0.5 - 5 GeV and the angle range (-5) - 3 mrad. In order to reflect the physical parameter range observed on the screens, the angle value was adjusted for each shot, based on the beam profile measurement on L1. Figure 10 presents the results of calibration performed on the raw data in Figure 10. We can observe the lineouts from the L3. The choice was made to use the data from the third screen since the energy resolution is higher. After running the algorithm, we found peak energies at E=[2.9 GeV, 3.1 GeV, 2.7 GeV, 3.3 GeV, 2.2 GeV] and a corresponding energy spread of [17%, 13%, 12.5%, 10%, 39.5%]. The electron beam parameters fluctuated due to the inherent instability of the acceleration but also due to laser conditions. The beam pointing and the beam size fluctuations are useful to examine the validity of our method.

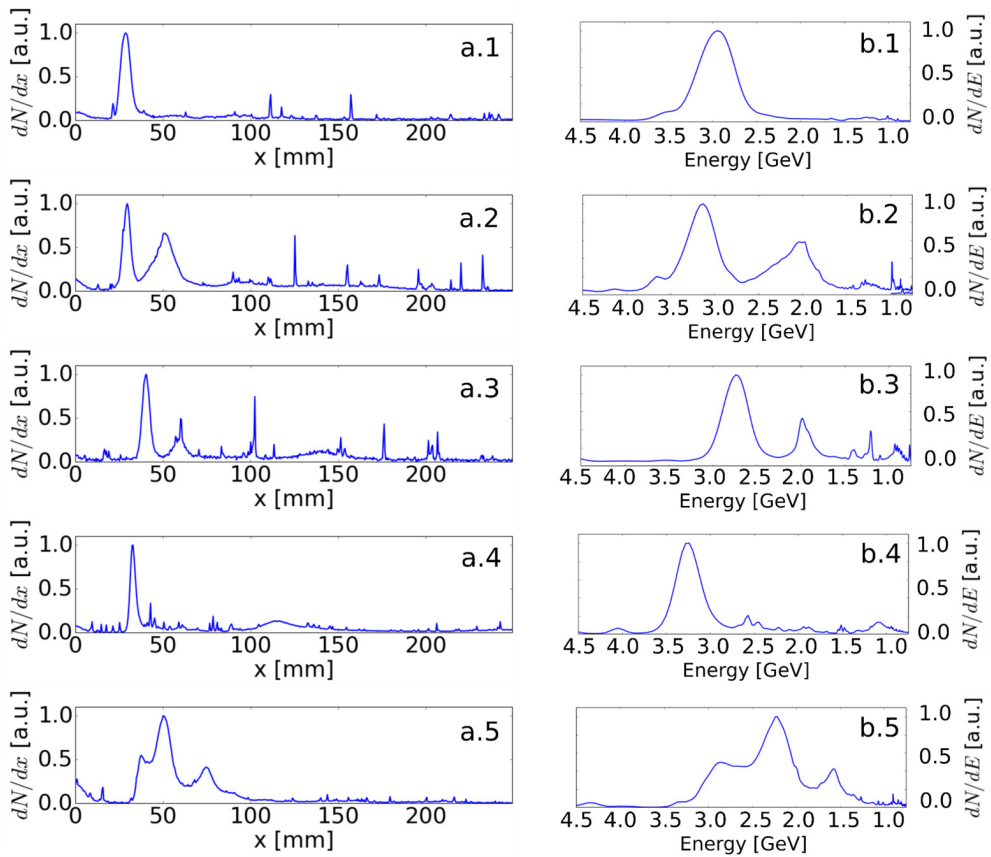

Figure 10 Lineouts from L3, before (a.1-a.5) and after (b.1-b.5) the calibration. The profiles are normalized to the maximum count of each acquired image.

We note that the algorithm could be further improved in order to fully automate the analysis. By itself this is a 2D optimization problem (angle and energy) and the search domain could be scanned with different methods. The computational cost and optimization would become more important if it is to be used in a high repetition environment. Furthermore, in order to completely automate the analysis, an automated image processing program can be useful in extracting information about the size of the beam profile and spectrum of an electron beam. This combination of detection and calibration can then be applied to the case of a high repetition laser wakefield acceleration setup for the accurate real-time optimization of beam energy and energy spread. The algorithm can be extended to 3D in order to account second order effects of the magnetic field.

## 5. Conclusions

We presented a new method suitable for accurately measuring and calibrating the spectrum of electron beams with pointing fluctuations generated from laser wakefield accelerators. The method was based on a combination of multiple screens, a dipole magnet spectrometer, and a reconstruction code the uses an algorithm to find the best trajectory that maximizes the accuracy (or minimizes the error). We evaluated the accuracy of the algorithm by comparing it with GEANT4 and the analytical formulae for the gyroradius. We found that the error between our code and the other tools was well below 1%, making it suitable for the analysis of accelerated GeV electron beams. We also reconstructed the energy and angle of simulated trajectories for a range of scenarios, and the algorithm works with a good accuracy below 1% in energy and 1% in angle. Furthermore, our method was used to calibrate experimentally obtained spectra from LWFA experiments performed at CoReLS, showing electron energy above 3 GeV. The combination of the measurement setup and the code presented in this work provided accurate diagnostics for a reliable characterization of electron beams produced from LWFA experiments. Full automation of this process would be beneficial for real-time optimization of electron energy, especially when using a high–repetition laser driver.